\documentstyle[prl,aps]{revtex}
\input{epsf}
\begin{document}
	
\draft

\twocolumn[\hsize\textwidth\columnwidth\hsize\csname 
@twocolumnfalse\endcsname

\title{Vortex Loop Phase Transitions in Liquid Helium, Cosmic Strings, 
and High-T$_{c}$ Superconductors}

\author{Gary A. Williams}
\address{Department of Physics and Astronomy, University of California,
Los Angeles, CA 90095}
\date{\today}
\maketitle
\begin{abstract}
The distribution of thermally excited vortex loops 
near a superfluid phase transition is calculated from a 
renormalized theory. The number density of loops with a given perimeter is 
found to change from exponential decay with 
increasing perimeter to algebraic decay as T$_{c}$ is approached, in agreement 
with recent simulations of both cosmic strings and high-T$_{c}$ 
superconductors. Predictions of the value of the
exponent of the algebraic decay at T$_{c}$ and of critical behavior
in the vortex density 
are confirmed by the simulations, giving strong support to 
the vortex-folding model proposed by Shenoy.

\end{abstract}

\vskip2pc]

The role of thermally excited vortex loops in three-dimensional
phase transitions where a U(1) symmetry is broken recently has become 
a prime topic in cosmology\cite{antunes,bett}
and high-T$_{c}$ superconductivity\cite{hightc,sudbo,ryu}.  These transitions 
are in the same universality class as 
the superfluid $\lambda$-transition in $^4$He.  In the helium case 
our original renormalization theory based on vortex 
loops\cite{williams,shenoy} 
has been extended to calculations of the specific heat\cite{spheat}
and to the dynamics of the transition\cite{dynamics}.  In this theory 
the Landau-Ginzburg-Wilson Hamiltonian is rewritten to cast it in terms 
of its elementary excitations, spin waves and vortex loops, providing an 
alternative method for carrying out the renormalization process, 
compared to the more traditional perturbation theories which expand 
the Hamiltonian expressed in terms of the order parameter.      

Here we further employ the loop theory to gain insight into recent 
simulations of the high-T$_{c}$ transition\cite{sudbo,ryu} and of 
cosmic-string phase transitions in the early universe\cite{antunes,bett}.  
The probability of 
occurrence of a vortex loop with perimeter $P$ is calculated, and in agreement 
with the simulations we find a crossover from quasi-exponential decay of the 
probability with perimeter at low temperatures, to purely algebraic decay 
precisely at T$_{c}\,$.  This provides strong support for the 
phenomenological ''Flory scaling'' treatment of the random-walking loops 
developed by Shenoy and co-workers\cite{chattopadhyay}.

In the vortex-loop theory the superfluid density $\rho_{s}$ is reduced by 
thermally excited loops whose average diameter $a$ increases as the 
temperature is increased, and the density is finally driven 
to zero at T$_{c}$ by loops of 
infinite size\cite{williams}.  Defining a dimensionless superfluid density by 
K$_{r}=\hbar^{2} \rho_{s} a_{o}$/$m^{2}k_{B}T$, where m is the 
mass of the $^4$He atom and $a_{o}$ the smallest ring diameter, the equation 
for the renormalized density is given by\cite{spheat}

\begin{equation}
{1 \over {K_r}}={1 \over {K_o}}+A_o\,\int_{a_o}^a  
{\;\left( {{a \over {a_o}}} \right)^6}\exp \left( {-{U(a) \over {k_BT}}} 
\right)\;{{da} \over {a_o}}
\quad  .
\end{equation}
Here A$_{o}$ = 4$\pi^{3}$/3, K$_{o}$ is the ''bare'' superfluid density 
resulting from the spin waves (and is the initial value 
of K$_{r}$), and 
U(a) is the renormalized energy of a ring, given by
\begin{equation} 
U(a)/k_{B}T=\pi ^2\int_{a_o}^a {\;K_r}(\ln \left( {{a \over {a_c}}} \right)+1)
{{da} \over {a_o}}+\pi ^2K_oC
\end{equation}
where C is a nonuniversal constant characterizing the core energy.  
For helium C and $a_{o}$ are determined from two experimental 
inputs, T$_c$ = 2.172 K and the amplitude of the superfluid 
density\cite{spheat}, yielding C = 1.03 and $a_{o}$ = 2.5 \AA .  
The effective core size $a_{c}$ in Eq.~(2) was suggested by Shenoy and 
co-workers\cite{shenoy,chattopadhyay} to be a result of the random walk of 
the loop giving rise to radial fluctuations of order $a_{c}$ about the average 
diameter.  This folding of the loop occurs because antiparallel vortex segments 
lower the energy.  A simple polymer-type 
calculation\cite{chattopadhyay} using energy-entropy arguments yields
$a_{c} / a = (K_{r} a / a_{o})^{\theta}$, where $\theta = d/(d+2)$ = 
0.6 in $d$ = 3 dimensions has the same form as the well-known Flory exponent 
of the self-avoiding walk.

Eqs.~(1) and (2) then constitute a coupled set of integral equations 
for the renormalized superfluid density, and can be solved recursively 
starting from the bare scale $a_{o}$ and iterating to distances 
greater than the correlation length $\xi = a_{o} / K_{r}$.  In 
practice these are converted to a set of coupled differential 
equations similar to the Kosterlitz recursion 
relations\cite{kosterlitz} for the 
two-dimensional case, and are solved using a Runge-Kutta technique
\cite{dynamics}.  As 
T is increased (K$_{o}$ decreased) the solution for $\rho_{s}$ falls to 
zero as (T$_{c}$-T)$^{\,\nu}$, with $\nu$ = 0.6717 for $\theta$ = 0.6.  This 
can be better matched to the most precise experimental value\cite{ahlers} 
$\nu$ = 0.6705 by adjusting to $\theta$ = 0.594, which is reasonable since 
it is known that the Flory-type arguments are not exact in three 
dimensions\cite{flory}. 

The arguments of Ref. 10 also yield a result for the average 
perimeter of a loop of diameter $a$,
\begin{equation}
{P \over {a_{o}}}=B\;\left( {{a \over {a_{o}}}} 
\right)^{{1 \mathord{\left/ {\vphantom {1 \delta }} \right. 
\kern-\nulldelimiterspace} \delta }}
\end{equation}
where B is a constant and the exponent $\delta$ = 2/($D$+2) = 0.4.  
This form for the perimeter was at least partially verified in 
the computer simulations of Ref. 10, and in the XY model simulations 
of Epiney\cite{epiney}.  For $P/a_{o}$ greater than about 20 the Epiney 
data for the average loop size versus average perimeter can  
be fit by Eq.~(3) with B = 1.8, although the resolution is 
poor because of scatter in the data resulting from the relatively small 
lattice (16$^{3}$) that was simulated.

The distribution of the density of loops with a given perimeter $P$ can 
be obtained from the theory outlined above, which is of interest 
because these distributions have now been measured in the 
cosmic-string\cite{bett} and high-T$_{c}$\cite{sudbo,ryu} simulations. 
The probability per unit volume for finding a loop 
of mean diameter between $a$ and $a$+d$a$ is
\begin{equation}
{\pi  \over {2\,a_o^3}}\;\left( {{a \over {a_o}}} \right)^2\;
\exp \left( {-{U(a) \over {k_BT}}} \right)\;{{da} \over {a_o}}
\quad ,
\end{equation}
Equating this to the probability 
D$(P)\,\,{{dP} \mathord{\left/ {\vphantom {{dP} {a_o}}} \right. 
\kern-\nulldelimiterspace} {a_o}}$
of finding the corresponding loop of perimeter 
between $P$ and $P$+d$P$ gives the probability distribution 
\begin{equation}
D(P)\;={\pi  \over {2\,a_{o}^3}}{\delta  \over B}\left( {{a \over 
{a_{o}}}} 
\right)^{{{3\delta -1} \over \delta }}\exp \left( {-{U(a) \over {k_BT}}} 
\right)\quad .
\end{equation}
For temperatures well below T$_{c}$\,, 
U($a$) $\sim a\ln a$, and hence D($P$) 
decreases exponentially with $P^{\delta}$.  Near T$_{c}$\,, however, the 
behavior is quite different.  By differentiating Eq.~(1) with respect to $a$ 
and substituting for the exponential term in Eq.~(5) gives
\begin{equation}
D(P)\;={\pi  \over {2\,a_o^3}}{\delta  \over B}\left( {{a \over {a_o}}} 
\right)^{-{{3\delta +1} \over \delta }}\left( {{{\partial \,
\left( {{1 \mathord{\left/ {\vphantom {1 {K_r}}} \right. 
\kern-\nulldelimiterspace} {K_r}}} \right)} \over {\partial \,a}}} 
\right) \quad .
\end{equation}
Precisely at T$_{c}$\,, K$_{r}$ from Eq. (1)  
satisfies the condition K$_{r} a / a_{o}$ = D$_{o}$ = 0.39, at least for 
values of $a$ greater than about 5$a_{o}$, and where D$_{o}$ is a 
universal constant\cite{spheat,pollack}.  This 
condition is the three-dimensional equivalent 
of the univeral ''jump'' of the superfluid density in two 
dimensions\cite{kosterlitz}.  
Inserting this result into Eq.~(6) and 
employing Eq.~(3) yields the prediction that at T$_{c}$ the loop 
distribution will cross over from exponential to algebraic decay with 
$P$\,,
\begin{equation}
D(P)\;={{\pi \,\delta \,B^{3\delta }} \over {2\,a_o^3A_oD_{o}}}\;
\left( {{P \over {a_o}}} \right)^{-\gamma }\quad ,
\end{equation}
where the exponent $\gamma$ = 3$\,\delta$ + 1.  For the ''Flory'' value 
$\delta$ = 0.4 this would predict $\gamma$ = 2.2\,.  This form for 
D($P$) signals the onset of loops of infinite size, since they no 
longer have a vanishing probability.  It should be 
noted that the algebraic decay is a consequence of the strong 
renormalization at T$_{c}$\,, where the screening of large loops by 
smaller ones causes the variation of U(a) to change\cite{dynamics} from 
$\sim a \ln a$ to $\ln a$ at T$_{c}$\,, causing the change from Eq.~(5) 
to Eq.~(7).  

The crossover from exponential to algebraic decay is a central feature 
observed in the recent cosmic-string\cite{bett} 
and high-T$_{c}$\cite{sudbo,ryu} 
simulations using lattices larger than 96$^{3}$, 
and which was also seen with more limited resolution in the 
earlier XY model simulations of 
Epiney\cite{epiney}.  The results of Antunes 
and Bettencourt\cite{bett} yield $\gamma$ = 2.23 $\pm$ 0.04 at 
T$_{c}$\,, which from the analysis 
above gives $\delta$ = 0.41 $\pm$ 0.01, 
in very good agreement with Shenoy's Flory-scaling prediction.  
Fits to the high-T$_{c}$ simulations of Nguyen and Sudbo\cite{sudbo} 
(in the zero-field, isotropic limit 
of their Villain model) give $\gamma$ = 
2.4\cite{fit}, leading to a higher 
value $\delta$ = 0.48.  However, the results 
of Ref.~2 show that $\gamma$ increases rapidly above T$_{c}$ to the 
value of 2.5 found by Vachaspati and Vilenkin\cite{vilenkin}, 
and hence an accurate determination requires bracketing temperatures very 
close to T$_{c}$. It is interesting that the result
$\gamma$ = 3$\,\delta$ + 1 apparently remains valid even above T$_{c}$\,, 
since the Vachaspati-Vilenkin value of $\gamma$ = 2.5 is based on the 
''Brownian'' value $\delta$ = 0.5.  

The magnitude of the loop distribution D($P$) at T$_{c}$ as 
calculated from either Eq.~(5) or~(7) appears to be about a factor of 3 
smaller than found in the simulations.  
Comparing the continuum calculation to 
the lattice results is made difficult by uncertainties in matching at 
the scale of the lattice spacing $a_{l}$;  in computing the magnitude 
of Eq.~(7) for the comparison the geometric value $a_{o} = \sqrt 2\,a_{l}$ 
was employed, but it is not entirely clear that this is the correct choice.

The total length per unit volume $\rho_{v}$ of the vortex loops can be found 
by multiplying D($P$) by $P$ and integrating. At T$_{c}$ this can be 
found explicitly using Eq.~(7),   
\begin{equation}
\rho _v(T_c)\;={{\pi \,\delta \,B } 
\over {2\,a_{o}^2A_{o}D_{o}\,
(3\delta -1)}}\;
\quad  .
\end{equation}
The quantity $\rho_{v}$(T$_{c}) a_{l}^{2}$ was postulated in Ref.~1 to 
be universal, with a value of 0.6 in lattice units (0.2 per 
placquette, with three placquettes per unit volume).  This means that 
the coefficient B in Eq.~(8) characterizing the  relationship between 
the perimeter and the average loop diameter 
must be universal, since all of the other parameters are.  As with 
D($P$) above, the magnitude of Eq.~(8) must be multiplied by a factor 
of about 3 to match with the lattice results.  For liquid helium, the 
vortex density\cite{Zph} at T$_{c}$ is then about 1 / $a_{o}^{2} 
\sim$ 1 x 10$^{15}$ /cm$^{2}$, which is many orders of magnitude 
higher than previous estimates\cite{kibble} which did not use 
a renormalized theory.

\begin{figure}
\epsffile[0 0 230 173]{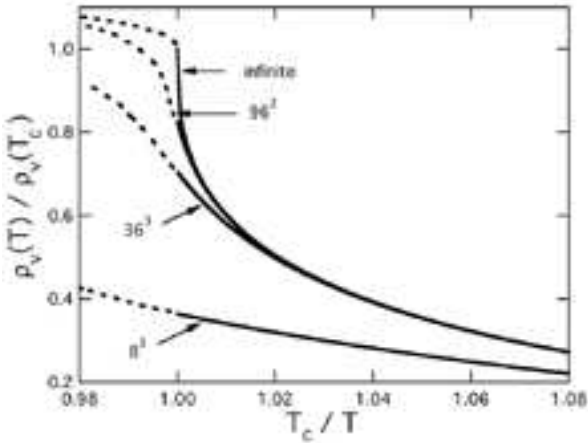}
\caption{Normalized vortex density as a function of T$_{c}$ / T, for 
several different lattice sizes.}
\label{}
\end{figure}

A further prediction of the loop theory is the existence of  
critical behavior in the vortex density 
at T$_{c}$\cite{Zph}, which has now been 
seen in the cosmic-string simulations\cite{antunes}.  This is 
shown in Fig.~1, which plots the normalized density versus T$_{c}$ / T, 
calculated by integrating Eq.~(5) very near the transition.  
The behavior above T$_{c}$ (dashed lines) is 
only conjectured, as the theory is not valid 
there.  To suppress the exponential 
variation arising from thermal excitation 
of the smallest loops, the core energy constant was taken to be the 
relatively large value C = 4/3, corresponding to the Villain model.  
The density just below T$_{c}$ is found 
to decrease from its value at T$_{c}$ as 
-(T$_{c}$-T)$^{\delta}$.  This exponent was not measured in the 
cosmic-string simulations, 
probably because of the relatively low core energy in that model, 
which would make it difficult to separate the algebraic 
behavior from the exponential.  However, the exponent of the density in the 
region just above T$_{c}$ was measured in Ref.~1 to be 0.39 $\pm$ 0.01, 
which is quite 
consistent with the value of $\delta$ = 0.41 determined above from the same 
type of simulation.  It is well known from scaling and renormalization-group 
studies that many critical exponents are identical above and below the 
transition, and it is likely that this exponent constitutes 
a further measurement of $\delta$ .  Since $\delta$ is less than one,
T$_{c}$ marks an inflection point in the density, where the curvature
changes sign.

Also shown in Fig.~1 is the effect of a finite-size lattice on the 
density near T$_{c}$\,.  For this the recursion relations are stopped at a 
finite scale that is a fraction $\beta$ = 0.75 smaller than the 
lattice size, where $\beta$ was found in Ref.~8 by comparing to the 
helium simulations of Ref.~15.  The effect of finite size is to smear 
out the critical behavior, and this explains why it was not apparent
in the early simulations\cite{shrock}, which only used a 
maximum 10$^{3}$ lattice size.

The coherence length in the loop theory can be identified with the 
diameter of the largest loop that is thermally excited\cite{williams}.
Since this is divergent at T$_{c}$\,, in a system of finite (but 
macroscopic) size the transition can be identified with the point 
where a loop just touches two opposing boundaries of the system.  
This is the percolation threshold for establishment of the infinite vortex 
cluster, as first proposed by Onsager 50 years ago, and verified in 
simulations\cite{onsager}.  At higher temperature even larger loops 
can be excited; these can take the form of a single vortex line 
passing from one side of the system to the other, with the topological 
return path of the loop being around the outside of the system.  These 
are known as ''string'' excitations in 
cosmology\cite{antunes,bett,vilenkin}.  The left-over strings from 
the rapid cooling through the phase transitions in the early universe 
are thought to be the source of matter\cite{kibble,zurek}.  
A laboratory example of this may be the observation of a few remnant vortex lines
\cite{schwarz} in a finite-sized sample of liquid $^4$He which has not been 
rotated or stirred, but simply cooled through the $\lambda$-transition.        


The key role of thermally excited vorticity in the present 
model of the phase transition 
allows a rather different viewpoint of the Kibble-Zurek 
mechanism\cite{kibble,zurek} for defect
formation in rapidly quenched transitions as discussed above.  This is of 
interest because of the possibility of carrying out such experiments in liquid 
helium\cite{kibble,zurek,mcclintock,volovik}.  In this view the vortices 
''created'' in a quench of superfluid $^4$He through T$_{c}$ are simply a 
perturbation on the equilibrium vortex density.  This perturbation is a 
consequence of the dynamics of large loops, which become slow in their 
response to external fields, and which are actually the 
source of the critical slowing down of the equilibrium transition
\cite{dynamics}.  In a rapid quench the large loops in the vicinity of 
T$_{c}$ cannot keep up and fall out of equilibrium, forming an excess density 
that survives to lower temperatures, finally decaying to the equilibrium line 
density only after the quench is stopped.  It may be possible to model this process 
analytically using the Fokker-Planck equation for the loop distribution 
function formulated in the dynamic theory of Ref.~9, with the temperature 
being a ramp function in time.  It should be noted that the exponent $\delta$ 
from above plays an important role in the vortex dynamics, since the 
frictional drag force on a loop is proportional to its total 
perimeter.  It was found in Ref.~9 that the dynamic exponent 
characterizing critical slowing down is given by x = z $\nu$, where 
z = (1-$\delta$)/$\delta$.  For the ''Flory'' value $\delta$ = 2/(d+2) this 
gives the scaling result z = d/2 = 3/2; for $\delta$ = 0.41 as found 
above, z = 1.44, a few percent smaller.  The possibility of 
deviations from the scaling 
result has been suggested previously in perturbative 
dynamic theories\cite{peliti}.

Vortex creation has been observed in superfluid $^3$He\cite{volovik} where the 
quench is induced by absorbed neutrons depositing their energy in a small 
region of the liquid, which heats it above T$_{c}$, and which is then
rapidly cooled back down by the surrounding cold liquid\cite{volovik}.  Although 
$^3$He is a p-wave BCS superfluid, vortex loops will still be 
associated with the phase transition as above, but due to the Ginzburg
criterion\cite{ginzburg} they will only be 
appreciable in a very narrow temperature range near T$_{c}$, since 
the zero-temperature coherence length of $^3$He is of order 500 \AA , 
compared with $a_{o}$ = 2.5 \AA \, in $^4$He. The theory\cite{volovik} of the 
quenched $^3$He involves the exponents $\delta$ and $\gamma$ at T$_{c}$; 
the use of 
the Brownian values\cite{vilenkin} $\gamma$ = 2.5 and $\delta$ =0.5 
needs to be reexamined in light of the present results.               

The vortex loop model also allows insights into the high-T$_{c}$ 
superconducting transition in zero field.  If T$_{c}$ is the point 
where the loops of infinite size act to bring all supercurrents to a 
halt, then it is {\it not} to be identified as 
the point where thermal de-excitation
of Cooper pairs is complete.  The continued existence 
of pairs above T$_{c}$ has 
been suggested in experiments, commonly known as the pseudogap 
phenomenon\cite{kivelson}.  The vortex loops constitute a concrete 
physical picture of the ''phase fluctuations'' postulated to give rise 
to this effect in Ref.~28.
The Cooper pairs above T$_{c}$ will not be the same as those 
below, since they will no longer be part of a macroscopic BCS-type 
condensate, which is destroyed by the vortices.  Presumably the pairs 
would be more localized excitations, on the scale of the 10-15 \AA 
$\;$ zero-temperature coherence length.

The vortices also offer a simple explanation\cite{hightc1} 
of the magnetic 
flux noise in YBCO samples that is observed 
to increase rapidly by many orders 
of magnitude over a temperature range of order 5 K near 
T$_{c}$\cite{clarke}.  When the loops 
being thermally excited terminate on the sample 
surface, they will induce dipolar 
current patterns on the surface, and this 
will generate magnetic flux that can 
be sensed by a detection loop above the 
surface.  As T$_{c}$ is approached from 
below both the number and size of the 
loops will increase, increasing the flux noise through the detector.  
This effect can also be observed in a low-T$_{c}$ 
superconductor, since the same vortex-loop transition 
occurs also in that case, but with the considerable difference that 
the large zero-temperature coherence length (several thousand \AA)
causes the temperature range where the vortices are appreciable to be very 
much closer to T$_{c}$.  The experiments\cite{clarke1} showed that 
indeed a very sharp flux-noise peak could be observed in a lead sample 
only within about 2 
mK of T$_{c}$, and that this was only an upper limit to the width
due to the resolution of the thermometry and the  
additional broadening that would be caused by a distribution of 
T$_{c}$'s across the sample.

In summary, a vortex-loop theory is able to provide physical insight 
into recent models of cosmic strings and high-T$_{c}$ 
superconductors.  The theory relates the critical exponents measured 
in the simulations to the random-walk nature of the loops, and the 
good agreement between the predicted and measured exponents provides 
strong support for the Flory-scaling ansatz of Shenoy\cite{shenoy}. 

Useful correspondence with L. Bettencourt, J. Clarke, M. Hindmarsh, 
A. Sudbo, and G. Volovik is gratefully acknowledged.
This work is supported by the U.S. National Science Foundation, grants 
DMR 95-00653  and  DMR 97-31523.


\begin{references}

\bibitem{antunes}
      N. Antunes, L. Bettencourt, and M. Hindmarsh,
	 {\it Phys. Rev. Lett.} 
	 {\bf 80}, 908 (1998).
\bibitem{bett}
      N. Antunes and L. Bettencourt,
	  {\it Phys. Rev. Lett.} 
	 {\bf 81}, 3083 (1998).
\bibitem{hightc}
      B. Chattopadhyay and S. R. Shenoy,
	 {\it Phys. Rev. B}  
	 {\bf 51}, 9129 (1995);
	 D. Dominguez {\it et al.},
	 {\it Phys. Rev. Lett.} 
	 {\bf 75}, 717 (1995);
	 M. Kiometzis, H. Kleinert, and A. Schakel,
	 {\it Phys. Rev. Lett.} 
	 {\bf 73}, 1975 (1994);
	 Z. Tesanovic, preprint, cond-mat/9801306.
\bibitem{sudbo}
      A. Nguyen and A. Sudbo,
	 {\it Phys. Rev. B}  
	 {\bf 57}, 3123 (1998);
	 {\bf 58}, 2802 (1998).
\bibitem{ryu}
     S. Ryu and D. Stroud,
	 {\it Phys. Rev. B} 
	 {\bf 57}, 14476 (1998);
\bibitem{williams} G. A. Williams, 
     {\it Phys. Rev. Lett.} {\bf 59}, 1926 (1987);
\bibitem{shenoy}
	 S. R. Shenoy, 
	 {\it Phys. Rev. B} {\bf 40}, 5056 (1989).
\bibitem{spheat} 
     G. A. Williams, 
     {\it J. Low Temp. Phys.} 
	 {\bf 101}, 421 (1995).
\bibitem{dynamics} G. A. Williams,
      {\it Phys. Rev. Lett.} {\bf 71}, 392 (1993);
     {\it J. Low Temp. Phys.} 
	 {\bf 93}, 1079 (1993).
\bibitem{chattopadhyay}
     B. Chattopadhyay, M. Mahato, and S.R. Shenoy,
	 {\it Phys. Rev. B} 
	 {\bf 47}, 15159 (1993).
\bibitem{kosterlitz}
     D. Nelson and J. M. Kosterlitz,
	 {\it Phys. Rev. Lett.} 
	 {\bf 39}, 1201 (1977).
\bibitem{ahlers}
     L. Goldner and G. Ahlers,
	 {\it Phys. Rev. B} 
	 {\bf 45}, 13129 (1992).
\bibitem{flory}
     N. Madras and A. Sokal, 
	 {\it J. Stat. Phys.}
	 {\bf 50}, 109 (1988);
\bibitem{epiney} 
     J. Epiney, 
	 Diploma thesis, ETH Zurich, 1990 (unpublished).
\bibitem{pollack}
     The quantity D$_{o}$/$\beta$ = 0.52 is the universal number which  
     should be compared with the value 0.49 found in the finite-size 
	 simulations of E. Pollack and K. Runge,
	 {\it Phys. Rev. B} 
	 {\bf 46}, 3535 (1992).
\bibitem{fit}
     The authors of Ref.~4 originally claimed the value of the
	 exponent to be 3, but they have now reported (private 
	 communication) that a value of 2.4 provides a better description of 
	 their data.
\bibitem{vilenkin} 
     T. Vachaspati and A. Vilenkin, 
	 {\it Phys. Rev. D}
	 {\bf 30}, 2036 (1984).
\bibitem{Zph}
	 G. A. Williams, 
	 {\it Z. Phys. B} 
	 {\bf 98}, 341 (1995).
\bibitem{kibble} 
     A. Gill and T. Kibble, 
	 {\it J. Phys. A} {\bf 29}, 4289 (1996).

\bibitem{shrock} G. Kohring, R. Shrock, and P. Wills, 
     {\it Phys. Rev. Lett.} 
	 {\bf 57}, 1358 (1986).
\bibitem{onsager} 
      L. Onsager,
      Nuovo Cimento Suppl. 
      {\bf 6} 249 (1949);
      J. Akao,
      {\it Phys. Rev. E} 
      {\bf 53}, 6048 (1996).
\bibitem{zurek}
     W. Zurek,
	 {\it Phys. Rep.} {\bf 276}, 177 (1996).
\bibitem{schwarz}
     D. Awaschalom and K. Schwarz
     {\it Phys. Rev. Lett.}
	 {\bf 52}, 49 (1984).
\bibitem{mcclintock}
     M. Dodd {\it et al.}
     {\it Phys. Rev. Lett.} 
     {\bf 81}, 3703 (1998).

\bibitem{volovik}
     G. Volovik,
	 {\it Czech. J. Phys.}
	 {\bf 46}, Suppl. S6, 3048 (1996); 
     V. Ruutu {\it et al.},
     {\it Phys. Rev. Lett.} 
     {\bf 80}, 1465 (1998);
     C. B\"auerle {\it et al.},
     Nature {\bf 382}, 332 (1996);
\bibitem{peliti}
     C. De Dominicus and L. Peliti,
     {\it Phys. Rev. B} 
	 {\bf 18}, 353 (1978).
\bibitem{ginzburg}
     V. Ginzburg,
     Sov. Phys. Solid State
     {\bf 2}, 1824 (1960).
\bibitem{kivelson}
     V. Emery and S. Kivelson,
	 {\it Nature}
	 {\bf 374}, 434 (1995).
\bibitem{hightc1}
     G. A. Williams,
	 {\it Physica B}
	 {\bf 194-196}, 1415 (1994).
\bibitem{clarke}
     M. Ferrari {\it et al.}
     {\it J. Low Temp. Phys.} 
	 {\bf 94}, 15 (1994).
\bibitem{clarke1}
     J. Clarke, private communication.
\end{references}
\end{document}